\pdfoutput=1

\documentclass{nime-alternate} 

\usepackage{anonymize}
\usepackage{hyperref}

\usepackage[T1]{fontenc}

\newcommand{\pnew}[1]{\textcolor{black}{#1}}

\usepackage{tikz}
\usepackage{pgfgantt}

\makeatother
\begin{document}

\usetikzlibrary{positioning, fit, calc, backgrounds, trees, arrows, shapes.geometric}

\tikzstyle{bela} = [rectangle,
rounded corners,
minimum width=2cm,
minimum height=1cm,
text width=2.5cm,
text centered,
draw=black,
thick,
fill=green!60,
font=\bfseries]

\tikzstyle{host} = [rectangle, rounded corners,
minimum width=2cm,
minimum height=1cm,
text width=2.5cm,
text centered,
draw=black,
text=white,
thick,
fill=blue!60,
font=\bfseries]

\tikzstyle{docker} = [rectangle, rounded corners,
minimum width=2cm,
minimum height=1cm,
text width=2.8cm,
text centered,
draw=black,
thick,
fill=cyan!60,
font=\bfseries]

\tikzstyle{lb} = [rectangle, rounded corners,
minimum width=0.1cm,
minimum height=0.1cm,
draw=black,
font=\bfseries,
fill=green!60,
text=white]

\tikzstyle{lh} = [rectangle, rounded corners,
minimum width=0.1cm,
minimum height=0.1cm,
draw=black,
fill=blue!60,
font=\bfseries,
text=white]

\tikzstyle{sensor} = [circle,
text width=0.5mm,
text centered,
draw=black,
thick,
fill=red!60,
font=\bfseries]

\tikzstyle{log} = [draw, thick, fill=black!20, text width=1cm, minimum width=0.6cm, align=center, font=\ttfamily]

\tikzstyle{tgt-bin} = [draw, thick, fill=yellow!60, align=center, font=\bfseries]

\tikzstyle{numpy} = [draw, thick, fill=pink!50, text width=1.2cm, align=center]
\tikzstyle{tflite} = [draw, thick, fill=olive!30, font=\ttfamily]

\tikzstyle{win} = [draw, thick, fill=red!50, align=center, minimum width=0.9cm, minimum height=0.7cm,font=\bfseries]

\tikzstyle{arrow} = [thick,->,>=stealth]
\tikzstyle{carrow} = [thick,->,>=stealth, dashed]

\conferenceinfo{NIME'23,}{31 May $-$ 2 June, 2023, Mexico City, Mexico.}

\title{Pipeline for recording datasets and running neural networks on the Bela embedded hardware platform}
\label{key}
\numberofauthors{4}
\author{
       \alignauthor
       \anonymize{Teresa Pelinski\titlenote{\anonymize{Equal contribution}}}\\
       \affaddr{\anonymize{Centre for Digital Music}}\\
       \affaddr{\anonymize{Queen Mary University of London, UK}}\\
       \email{\anonymize{t.pelinskiramos@qmul.ac.uk}}\\
       \alignauthor
       Rodrigo Diaz\raisebox{9pt}{$\ast$}\\
       \affaddr{\anonymize{Centre for Digital Music}}\\
       \affaddr{\anonymize{Queen Mary University of London, UK}}\\
       \email{\anonymize{r.diazfernandez@qmul.ac.uk}}\\
       \alignauthor
       \anonymize{Adán L. Benito Temprano}\\
       \affaddr{\anonymize{Centre for Digital Music}}\\
       \affaddr{\anonymize{Queen Mary University of London, UK}}\\
       \email{\mbox{\anonymize{a.benitotemprano@qmul.ac.uk}}}\\
       \and
       \alignauthor
       \anonymize{Andrew McPherson}\\
       \affaddr{\anonymize{Dyson School of Design Engineering}}\\
       \affaddr{\anonymize{Imperial College London, UK}}\\
       \email{\mbox{\anonymize{andrew.mcpherson@imperial.ac.uk}}}\\
}

\date{30 July 1999}

\maketitle

\begin{abstract} 

       Deploying deep learning models on embedded devices is an arduous task: oftentimes, there exist no platform-specific instructions, and compilation times can be considerably large due to the limited computational resources available on-device. Moreover, many music-making applications demand real-time inference. Embedded hardware platforms for audio, such as Bela, offer an entry point for beginners into physical audio computing; however, the need for cross-\hspace{0pt}compilation environments and low-level software development tools for deploying embedded deep learning models imposes high entry barriers on non-expert users.

       We present a pipeline for deploying neural networks in the Bela embedded hardware platform. In our pipeline, we include a tool to record a multichannel dataset of sensor signals. Additionally, we provide a dockerised cross-compilation environment for faster compilation. With this pipeline, we aim to provide a template for programmers and makers to prototype and experiment with neural networks for real-time embedded \pnew{musical} applications.

\end{abstract}


\sloppy

\section{Introduction} 

There are many instruments in the NIME community based on embedded systems such as single-board computers (e.g. Bela, Raspberry Pi) or microcontrollers (e.g. Teensy, Arduino) \cite{Sullivan2020, Berdahl2014, Zayas-Garin2021, MeneseS2019}. Plenty of these platforms provide open-source APIs \pnew{and IDEs}  that abstract, among others, the complexities of \pnew{compilation or} interfacing with peripherals, which along with supportive online learning communities \cite{Kuznetsov2010}, \pnew{well-documented code bases and tutorials,} make them a valuable teaching resource for music and audio programming courses \cite{Masu2021a}.

\pnew{One of the most attractive features of these devices} to instrument designers is their self-containedness. Embedded hardware-based instruments can work off-the-shelf \pnew{and are less susceptible to issues caused by system updates, which may result in software and hardware compatibility problems when using general-purpose computers} \cite{Morreale2017a}. In general, they require less maintenance than laptop-based instruments \cite{Berdahl2014}, which, in turn, is a desirable feature in terms of the instrument's longevity \cite{Morreale2017}. Their small size also allows for integrating them into the instrument's body.

Many of these platforms provide APIs that simplify their usage. These APIs are typically written in C++, but many have incorporated other audio programming languages. In this context, Morreale et al. \cite{Morreale2017a} introduce the notion of ``pluggable communities'', which refers to how disparate communities establish a connection when users are empowered to map their knowledge into a new domain. Incidentally, there is a growing interest in deep learning techniques in NIME \cite{Martin2020a, Tahiroglu2021b, Bergomi2022, Pelinski2022a}. However, deploying neural models into embedded platforms is an arduous task: oftentimes, no platform-specific instructions exist, and the large building times (due to the limited computational resources) complicate debugging. Frequently, the APIs provided by these platforms do not integrate deep learning inference engines, and the programmer needs lower-level software development skills to compile them. In addition, the deep learning models need to be very computationally efficient to run in real-time. While there exist instances of porting deep learning frameworks into embedded systems in the context of NIME~\cite{Martin2020, DeviS2021, Stefani2022}, considerable effort is still needed to make these accessible for non-expert makers.

This paper presents a pipeline to run neural networks in real-time in the Bela\footnote{\url{https://bela.io/}} hardware platform \cite{McPherson2015}, a platform extensively used in NIME~\cite{Zayas-Garin2021,Moro2016,Morreale2017a, GonzalezSanchez2018}. Given the stringent real-time requirements of musical instruments and interfaces, we have chosen Bela due to its low input-output latency \cite{McPherson2016}. As part of our pipeline, we include a tool to record a multichannel dataset of sensor signals. Additionally, we provide a dockerised cross-compilation environment for faster compilation.  With this pipeline, we aim to \pnew{streamline} the prototyping and experimentation with simple deep learning models for real-time embedded audio applications, and to contribute towards bridging the deep learning for audio and the embedded hardware communities in NIME.

\section{Background}

\pnew{The practice of running neural networks in embedded devices is referred to by different names across various domains, such as \textit{edge AI} in networked devices and cloud applications, \textit{AIoT} in the context of artificial intelligence integrated into everyday devices, and \textit{tinyML} for devices that operate on a few milliwatts of power.}

In this paper, we continue with the terminology adopted on the NIME 2022 workshop \textit{Embedded AI for NIME: Challenges and Opportunities}~\cite{Pelinski2022a} and \pnew{use the term ``embedded AI'' with some nuances: first, we focus on embedded platforms that have low-power and low-resourced CPUs, but that allow for real-time sensor or audio signal processing; and secondly, we concentrate on deep learning\footnote{\pnew{Performing deep learning inference in a resource-constrained device might seem counter intuitive: the ``deep'' in ``deep learning'' implies that the model has a considerable number of layers and parameters, and consequently, that it will need significant computational resources to run. A more accurate term would be ``shallow'' learning, however, we use the term deep learning due to its extended adoption. }} rather than other techniques that fall under the umbrella of artificial intelligence.}

\subsection{\pnew{Embedded Inference}} \label{ssec:emb-inf}

Inferencing with deep learning models for real-time audio is a computationally demanding task, even for conventional computers. \pnew{For instance, to generate audio at 44.1kHz, the model should be able to generate at least 44100 samples every second. Since they are typically based on matrix operations, neural networks are usually trained and run in Graphics Processing Units (GPUs). There exist embedded computers with integrated GPUs\footnote{Such as the Nvidia Jetson Nano (\url{https://developer.nvidia.com/embedded/jetson-modules}) or the Coral Dev Board Mini (\url{https://coral.ai/products/dev-board-mini/}).}}. \pnew{However, the communication overhead of interfacing between CPU and GPU can be challenging for real-time audio applications, as well as the balance between meeting the sampling rate and staying within limits for latency and jitter~\cite{Renney2020}. Recent advancements have been made in this direction in the context of real-time audio effects~\cite{Skare2020}.} Non-GPU alternatives for deep learning exist: Kiefer~\cite{Kiefer2021} explores the usage of field-programmable gate arrays (FPGAs), which can run large parallel processes at very high frequencies, \pnew{though they are complex to manipulate when compared to commercial embedded computers such as Bela or Raspberry Pi \cite{Kiefer2021}}.

\pnew{A perhaps more straightforward alternative is to directly run the networks on the CPU. Mittal et al. \cite{Mittal2022} survey the field of deep learning in CPUs, and find that CPUs can outperform GPUs for large models and batch sizes due to their greater available memory \cite{Wang2019}. The memory advantage is also beneficial for networks where the number of computations rises with sequence lengths but parallelisation is complex due to sequential dependency (e.g. RNNs) \cite{Zhang2018}. Furthermore, in embedded systems, the CPU can be a better suited choice for inference than the GPU since continuous inference in the GPU can lead to a high energy consumption \cite{Zhang2018}. There exist hardware accelerators that can be attached to embedded computers to speed up training and inference\footnote{Such as Google Coral (\url{https://coral.ai/products/accelerator}) or the Intel Neural Compute Stick (\url{https://www.intel.com/content/www/us/en/developer/articles/tool/neural-compute-stick.html})}, however, the software must be able to leverage these optimisations \cite{Hanhirova2018}.}

\pnew{An alternative approach to hardware optimisation is to reduce the network size through compression techniques such as pruning, quantisation or knowledge distillation, among other strategies. Pruning techniques reduce the model's size by discarding a substantial amount of weights in a neural network without significantly decreasing its accuracy. Alternatively, quantisation strategies quantise the weights and activations of a network to a lower-precision datatype. Lastly, in knowledge distillation approaches, a small student model mimics a larger teacher model.}
\pnew{In the case of pruning, the ratio of pruned parameters to the number of parameters originally present in the network (sparsity) can be used as a proxy to estimate the performance of a network in a target platform. However, calculating the exact time a model will need to run, which is relevant when prototyping for real-time applications, is complex since many other factors influence the efficiency of a model, such as the enabled hardware optimisations, memory management and latency inherent to certain data operations (e.g. FFTs). }



\pnew{Finally, hard real-time systems are usually programmed with compiled languages such as C or C++, which favour deterministic code. Some deep learning frameworks provide C++ distributions (e.g. Libtorch\footnote{\url{https://pytorch.org/docs/stable/jit.html}} for PyTorch and TFLite\footnote{\url{https://www.tensorflow.org/lite}} for TensorFlow), however, they tend to rely on resizable data structures that can allocate memory dynamically (e.g. C++'s \texttt{std::vector.resize()}), which makes them potentially inappropriate for real-time implementations \cite{Chowdhury2021}. Stefani et al. \cite{Stefani2022} run a comparison of audio classification performance of the TFlite, Libtorch, ONNX Runtime\footnote{\url{https://onnxruntime.ai/}} and RTNeural \cite{Chowdhury2021} inference engines. The authors find that these frameworks can be used safely for hard real-time applications.
       Alternatively, the IREE (Intermediate Representation Execution Environment) is a compiler and runtime stack based on the MLIR (Multi-Level Intermediate Representation) compiler infrastructure \cite{Vasilache2022}, which converts the models into an intermediate representation that allows optimising the model for the target platform hardware. A recent Google Summer of Code project \cite{Pierce2022} developed support for compiling, running and benchmarking IREE projects on Bela.}


\subsection{\pnew{Existing Tools}}

Whilst these \pnew{technical} contributions represent significant steps toward embedding \pnew{deep learning} models for musical applications, there is still a need for tools that \pnew{streamline these achievements to facilitate experimentation and prototyping.} In their Cognitive Dimensions of Notations Framework, Blackwell and Green~\cite{Blackwell2003} include the ``viscosity'' dimension, which refers to the resistance of a system to change. In this sense, the process of compiling inference engines and prototyping with deep learning models for embedded platforms is viscous, since it often takes a considerable amount of attempts \pnew{and it involves a variety of programming languages, frameworks and devices}. Although Blackwell and Green use the term viscosity to refer to the number of actions needed to accomplish a goal, here we might extend it to the time it takes for a system to change. A pipeline for compiling and running deep learning models on Bela reduces the viscosity of the task by providing a \pnew{streamlined set of steps and templates, including a cross-compilation environment that reduces the compilation times.}

\pnew{The pipeline is intended for non-expert programmers with sufficient skills to follow a tutorial involving interacting with the target platform trough the CLI and coding in C++ and python. It aims to encourage prototyping and experimentation through code rather than interfaces.}
\pnew{There exist a few laptop- and interface-based tools (typically in Pd and Max/MSP, or as VST plugins) that allow applying deep learning and machine learning models in real-time to input audio or sensor signals, such as FluCoMa \cite{Tremblay2021}, the nn\_tilde\footnote{\url{https://github.com/acids-ircam/nn_tilde}} and torchplugins\footnote{\url{https://github.com/rodrigodzf/torchplugins}} Max/MSP and Pd externals, or the Neutone VST plugin\footnote{\url{https://neutone.space/}}; yet to the authors' knowledge, none specifically offers an embedded implementation.} However, there are a couple of examples of audio-based embedded models: the real-time neural audio synthesis model RAVE \cite{Caillon2021}, that has been embedded\footnote{\url{https://youtu.be/jAIRf4nGgYI}} into a Raspberry Pi and an Nvidia Jetson Nano,
and the Neurorack \cite{DeviS2021}, a Eurorack module running a neural source-filter model, also embedded on the Nvidia Jetson Nano.

Meanwhile, music-making \pnew{deep learning} models involving the performer's body and gesture have received much less attention, although the 2022 workshop \textit{Embodied Perspectives on Musical AI} held at the University of Oslo in Norway \cite{Erdem2022} manifested its increasing relevance. \pnew{There exist many laptop- and interface-based machine learning toolkits for gesture classification and mapping, such as GIMLeT (Max/MSP) \cite{Visi2021}, ml.lib (Max/MSP, Pd) \cite{Bullock2015}, XMM (C++, python and Max/MSP) \cite{Francoise2014} or MnM (Max/MSP) \cite{Bevilacqua2005}}, however, there are only a few embedded approaches related to \pnew{deep learning} and gesture in the context of NIME, such as Martin et al.'s work \cite{Martin2020}, where a recurrent neural network runs in a Raspberry Pi to predict the performer's control gestures. Whilst some of these projects provide very detailed instructions for deploying the embedded models, we aim to provide a more general and model-agnostic pipeline for practitioners to apply to their own projects.



\section{Pipeline}
\begin{figure}[hbt!]
       \centering
       \begin{tikzpicture}[node distance=2.8cm]
              \node (rec) [bela, align=center] {Dataset\\recording};
              \node (proc) [host, right= 1.8cm of rec, align=center] {Dataset\\processing};
              \node (train) [host, below=  1.3cm of proc, align=center] {Model\\training};
              \node (xc) [host, below= 1.1cm of train, align=center] {Bela source code cross-compilation};
              \node (exe) [bela, left= 1.8cm of xc, align=center] {Run\\executable};

              \draw [arrow] (rec) -- node[anchor=north, align=left, below=0.3cm of rec]{Raw data} (proc);
              \draw [arrow] (proc) -- node[anchor=west, right=0.8cm of proc, align=left] {Synchronised\\and processed\\dataset} (train);
              \draw [arrow] (train) -- node[anchor=west, right=0.8cm of train, align=left] {TFLite\\model} (xc);
              \draw [arrow] (xc) -- node[anchor=north, align=center, below=0.3cm of xc] {Bela-\\compatible\\executable}(exe);

              \draw [arrow] (rec) -- (proc);
              \draw [arrow] (proc) --  (train);
              \draw [arrow] (train) --  (xc);
              \draw [arrow] (xc) -- (exe);

              \node (log)[log]at ($(rec)!0.5!(proc)$) {.log};
              \node (tgt-bin)[tgt-bin]at ($(xc)!0.5!(exe)$) {exe};
              \node (numpy)[numpy]at ($(proc)!0.5!(train)$) {\texttt{numpy} array};
              \node (numpy)[numpy]at ($(proc)!0.5!(train)$) {\texttt{numpy} array};
              \node (tflite)[tflite]at ($(train)!0.5!(xc)$) {.tflite};

              \matrix [draw, thick,  rounded corners, above=0.6cm of log, fill=black, text=white, font=\bfseries]{
                     \node [lb,label=right:Bela] {}; &
                     \node [lh,label=right:Host] {};   \\
              };

       \end{tikzpicture}
       \caption{Pipeline for recording datasets and running neural networks on Bela. Green and blue nodes indicate, respectively, that the code runs in Bela or in the host machine.
       }\label{fig:pip}
\end{figure}
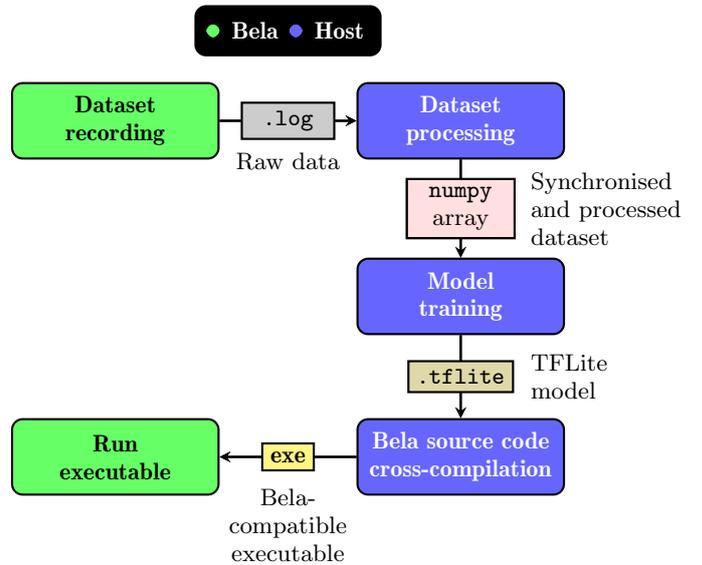

We present a pipeline to record a dataset of signals, export a light model trained on those signals, and run the model in real-time on Bela.
An overview of the pipeline is given in Figure~\ref{fig:pip}. It should be noted the pipeline \pnew{relies on} a host machine for dataset processing, training and exporting a light model, and cross-compiling the inference code. The pipeline's code and instructions for each step are available in Github:

\begin{center}
       \url{https://github.com/pelinski/bela-dl-pipeline}
\end{center}

In the pipeline's first stage, a dataset of sensor signals can be recorded in a single or in multiple Bela boards. The raw data files are then transferred into the host machine for the data processing step, where the signals recorded on different boards are aligned sample-wise, and the dataset is converted into a numpy\footnote{\url{https://numpy.org/}} array, which can be loaded into deep learning \pnew{python} frameworks such as Tensorflow or PyTorch. After the model is trained on the dataset, it must be exported as a \texttt{.tflite} file, since the Bela inference code is based on the TFLite C++ library. Finally, the inference code is cross-compiled and transferred to Bela, where it can be executed. Below, we describe each step in detail.

\subsection{Recording and processing datasets}

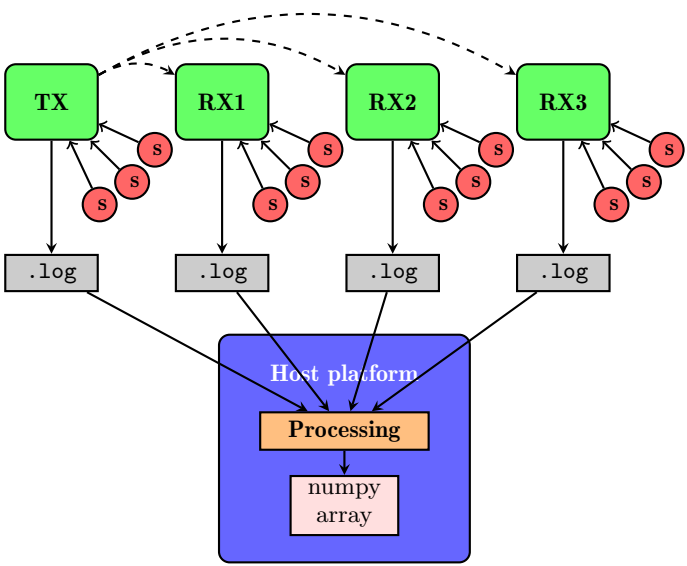
\begin{figure}[hbt!]
       \centering
       \begin{tikzpicture}[edge from parent/.style={draw, thick, <-, thick}]
              \node (tx)  [bela, text width=1cm, minimum width=0.6cm, align=center] {TX} [counterclockwise from=-65]
              child { node [sensor]  {s}}
              child { node [sensor]  {s}}
              child { node [sensor]  {s}}
              ;

              \node (rx1)  [bela, text width=1cm, minimum width=0.6cm, right=of tx, align=center] {RX1} [counterclockwise from=-65]
              child { node [sensor]  {s}}
              child { node [sensor]  {s}}
              child { node [sensor]  {s}}
              ;

              \node (rx2)  [bela, text width=1cm, minimum width=0.6cm, right=of rx1, align=center] {RX2} [counterclockwise from=-65]
              child { node [sensor]  {s}}
              child { node [sensor]  {s}}
              child { node [sensor]  {s}}
              ;

              \node (rx3)  [bela, text width=1cm, minimum width=0.6cm, right=of rx2, align=center] {RX3} [counterclockwise from=-65]
              child { node [sensor]  {s}}
              child { node [sensor]  {s}}
              child { node [sensor]  {s}}
              ;

              \node (log_tx)  [log, below= 1.5cm of tx] {.log};
              \foreach \i in {1,...,3} \node[log, below=1.5cm of rx\i] (log_rx\i) {.log};

              \draw[carrow] (tx) to [bend left] (rx1);
              \draw[carrow] (tx) to [bend left] (rx2);
              \draw[carrow] (tx) to [bend left] (rx3);

              \draw[arrow] (tx) -- (log_tx);
              \draw[arrow] (rx1) -- (log_rx1);
              \draw[arrow] (rx2) -- (log_rx2);
              \draw[arrow] (rx3) -- (log_rx3);

              \node (host) [host, below=0.55cm of log_tx, text depth=2cm, inner sep=0.4cm, font=\bfseries] at (current bounding box.south) {Host platform};

              \node (proc)  [draw, thick,  fill=orange!50, text width=2cm, below=of log_tx, minimum width=0.6cm, align=center, font=\bfseries] at ([yshift=4.6em]host.center) {Processing};

              \node (numpy)  [numpy, below=of proc] at ([yshift=2em]host.center) {numpy array};

              \draw [arrow] (log_tx) -- (proc);
              \draw [arrow] (log_rx1) -- (proc);
              \draw [arrow] (log_rx2) -- (proc);
              \draw [arrow] (log_rx3) -- (proc);
              \draw [arrow] (proc) -- (numpy);

       \end{tikzpicture}
       \caption{Example of recording a dataset on four different Bela \pnew{boards}, with three sensors (red circles) connected to each. The dashed line represents the clock signal (a digital bit) sent from the transmitter Bela (TX) to the receiver Bela boards (RX1, RX2, RX3). The generated log files are processed in the host platform. }\label{fig:data}
\end{figure}

In the first stage of the pipeline, a dataset from analog (e.g. piezo sensors, microphone signals) or digital (e.g. distance sensors, rotary encoders) inputs is recorded on Bela. Our code, based on the \texttt{BelaParallelComm} library, allows recording datasets simultaneously on various Bela boards, which enables capturing more channels than those available on a single Bela board (i.e. 8 analog and 16 digital inputs). As illustrated in Figure~\ref{fig:data}, one of the Bela boards acts as a transmitter (TX), \pnew{whilst} the other boards \pnew{act} as receivers (RXs). Every number of frames, the \pnew{TX Bela} sends a digital bit to the \pnew{RXs Bela boards}. Besides logging the sensor signals' values, \pnew{all Bela boards} log the frame at which the bit was sent (in the case of the TX), or received (in the case of the RXs).

In the second stage of the pipeline, these log files are transferred into the host machine, where the \texttt{DataSyncer} library synchronises the signals at sample level and returns a single numpy matrix in which every row corresponds to a sensor channel and every column corresponds to a timestep. This matrix can later be loaded into the user's preferred \pnew{python} deep learning framework. If the dataset is recorded only on one Bela, the \texttt{DataSyncer} library will simply convert it into a numpy matrix.

Recording datasets on multiple Bela \pnew{boards} is valuable for analysis tasks (e.g. to analyse the timing variations of various performers simultaneously playing the same piece on a piezo-based instrument) or if the Bela boards are used as an interface to stream sensor data to a model running in the host machine. \pnew{Inferencing with a deep learning model on Bela with inputs simultaneously proceeding from various boards is a complex process involving real-time communication within Bela boards, which would require sufficient bandwidth to enable the transmission of multichannel data to the Bela board executing the model, or alternatively, distributing the network across boards.}

\subsection{Training and exporting a model}
Given the limited computational capabilities of Bela, in the third stage of the pipeline, the model training is carried out on the laptop or, alternatively, on a \pnew{computing} cluster. In the provided example code, we use PyTorch, a widely extended \pnew{python} deep-learning framework. However, Stefani et al.~\cite{Stefani2022} find that the TFLite inference engine is faster than PyTorch's C++ inference engine. For this reason, we wrote the model in PyTorch and then exported it to Tensorflow Lite using the \texttt{TinyNeuralNetwork}\footnote{\url{https://github.com/alibaba/TinyNeuralNetwork}} library. \pnew{The process of converting PyTorch models to the TFLite format may not always be straightforward, particularly when PyTorch primitives lack a direct mapping to an equivalent TFLite primitive. In such cases, it may be necessary to prototype the model without using these primitives (i.e. explicitly writing the model equations). Alternatively, the model may be prototyped using Keras o Tensorflow, since} these frameworks can natively export \pnew{a TFLite-compatible} model.

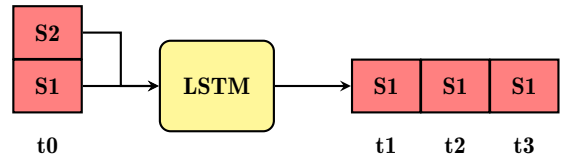
\begin{figure}[hbt!]
       \centering
       \begin{tikzpicture}[node distance=1cm]
              \node (in_1) [win] {S1};
              \node (in_2) [win, above=-0.02cm of in_1] {S2};

              \node (lstm) [draw, thick,  thick, rounded corners, fill=yellow!50, minimum width=1.5cm, minimum height=1.2cm,  right=of in_1, align=center, font=\bfseries] {LSTM};

              \node (out_1) [win, right=of lstm] {S1};
              \node (out_2) [win, right=-0.02cm of out_1] {S1};
              \node (out_3) [win, right=-0.02cm of out_2] {S1};
              \node (t0) [below= 0.2cm of in_1,font=\bfseries] {t0};
              \node (t1) [below= 0.2cm of out_1,font=\bfseries] {t1};
              \node (t2) [below= 0.2cm of out_2,font=\bfseries] {t2};
              \node (t3) [below= 0.2cm of out_3,font=\bfseries] {t3};

              \draw [arrow] (in_2.east) -- + (0.5,0) |-  (lstm.west);
              \draw [arrow] (in_1) --  (lstm.west);
              \draw [arrow] (lstm) --  (out_1);

       \end{tikzpicture}
       \caption{Example model. The model receives a window of the sensor channels S1 and S2 and predicts the following three windows of sensor S1.
       }\label{fig:mod}
\end{figure}

In the provided example, we train an LSTM network that receives two sensor signals (S1 and S2) and predicts the future values for the sensor S1 signal. This is illustrated in Figure~\ref{fig:mod}. A model that predicts the future behaviour of sensor signals has interesting options for performance, such as instruments that speculate about what the performer \pnew{might do, or instruments that sonify the prediction error.} In our setup, sensor S1 was an accelerometer attached to a pendulum, and sensor S2 was a piezo sensor attached to a drumstick. The dataset consists of recordings of the sensor signals when hitting the pendulum (S1) with the drumstick (S2). A window of 32 samples of the two sensor channels is passed into the LSTM, which predicts the next three windows of the S1 signal (in total, 96 samples). The model, which runs in real-time in Bela, had 7096 parameters and a size of 1MB when exported as \texttt{.tflite}. The next section will discuss the coding practices we implemented for this model to run in real-time.


\subsection{Cross-compiling}
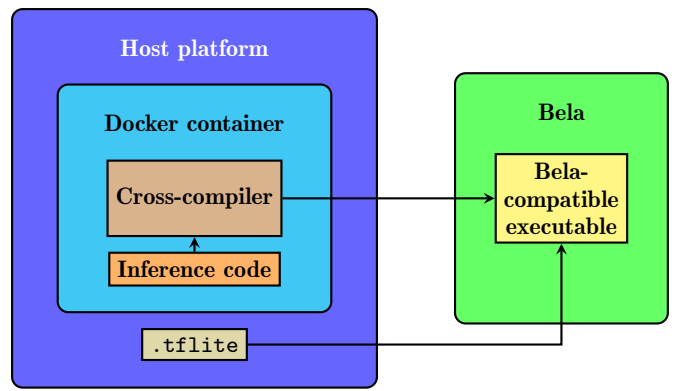
\begin{figure}[hbt!]
       \centering
       \begin{tikzpicture}
              \node (host) [host, text depth=4cm, inner sep=0.4cm, text width=4cm] {Host platform};
              \node (docker) [docker, text depth=2cm, inner sep=0.4cm] at (host.center){Docker container};
              \node (xc) [draw, thick, fill=brown!60, font=\bfseries, minimum height=1cm] at (docker.center){Cross-compiler};
              \node (sc) [draw, thick, fill=orange!60, font=\bfseries] at ([yshift=1.8em]docker.south){Inference code};
              \node (tflite) [tflite] at ([yshift=1.8em]host.south){.tflite};

              \node (bela) [bela, text depth = 2.3cm, right=of host, inner sep=0.4cm,text width=2cm] {Bela};,
              \node  (tgt-bin) [draw, thick, fill=yellow!60, align=center, font=\bfseries] at (bela.center){Bela-\\compatible\\executable};

              \draw [arrow] (xc) -- (tgt-bin);
              \draw [arrow] (sc) -- (xc);
              \draw [arrow] (tflite) -| (tgt-bin);
       \end{tikzpicture}
       \caption{Cross-compilation for Bela. The inference source code is compiled in a Docker container in the host machine (e.g. a laptop). Both the Bela-compatible executable and the \texttt{.tflite} model are then transferred to the Bela.}\label{fig:xc}
\end{figure}

Building a complex library or program can take considerable time on an embedded device, which makes prototyping and debugging tedious. Cross-compilation reduces building times by compiling the program in a host platform (e.g. a laptop) with greater computational resources. To facilitate cross-compilation for Bela, in the fourth stage of the pipeline, we provide a dockerised container to encapsulate the cross-compiler. The workflow is illustrated in Figure~\ref{fig:xc}. This \pnew{enables} compiling Bela code on any host that can run Docker\footnote{\url{https://www.docker.com/}} (i.e. Linux, Windows, MacOS).  Docker is a tool to package software and its dependencies in a container, which allows running that software across platforms without needing to install OS-specific dependencies each time. Inside the Docker container, we use CMake\footnote{\url{https://cmake.org/}} for cross-compiling. CMake is an open-source tool commonly used for building C++ projects, in which the instructions for compilation are passed in a \texttt{CMakeLists.txt} file. CMake also allows cross-compiling code by using a ``Toolchain'', a file that describes the target platform. The toolchain file for Bela is included in the provided Docker container.

To cross-compile Bela code using the Tensorflow library, the library must be previously built inside the container. Once the Tensorflow library has been compiled, its location can be passed to the compiler using CMake. In our repository, we provide the pre-compiled libraries for Bela (and instructions to cross-compile it), as well as example code and the \texttt{CMakeLists.txt} file needed to compile it.


\subsection{Inference in real-time and multi-threaded processing}


In this section, we discuss the real-time coding practices that should be \pnew{followed in} the inference code for the deep learning models to run in real-time (fifth stage of the pipeline). These practices apply to any real-time system, but they are particularly relevant here since neural networks' inference is a computationally expensive operation. Template code for running inference in Bela is included in the provided repository. It should be noted that these practices will enable the model to run in real-time only if the model is light enough to run using the device's CPU. For instance, Esling et al.~\cite{Esling2020} evaluate the theoretical embeddability of deep learning models in terms of compression and complexity according to three metrics: floating point operations, model disk size and number of read-write operations. Further work is needed to determine these parameters' thresholds on Bela.

\subsubsection{Pre-allocating memory}
Allocating memory is a non-deterministic process, which means that its duration can not be known in advance. In order to guarantee our code can run in real-time, we need to ensure that every part of the code has a bounded execution time,
and that it will meet our real-time deadlines. Therefore, memory needs to be allocated before the audio processing starts. In the Bela API, memory should be allocated in the function \texttt{setup()}, which runs at the beginning of a program's execution and before the audio processing starts.

\subsubsection{Multi-threading to avoid underruns}
\begin{figure*}[hbt!]
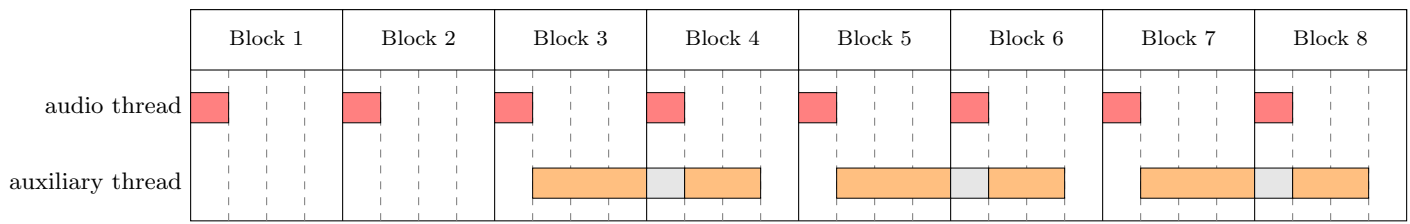

       \centering
       \hspace*{-1.5cm}\begin{ganttchart}[
                     vgrid={*3{gray, dashed}, *1{black}},
                     bar/.append style={fill=red!50, line width=0.2pt, align=right},
                     x unit=0.5cm,
                     title height=1,
                     y unit title=0.8cm]{1}{32}
              \gantttitle{Block 1}{4}
              \gantttitle{Block 2}{4}
              \gantttitle{Block 3}{4}
              \gantttitle{Block 4}{4}
              \gantttitle{Block 5}{4}
              \gantttitle{Block 6}{4}
              \gantttitle{Block 7}{4}
              \gantttitle{Block 8}{4}\\
              \ganttbar{audio thread}{1}{1}
              \ganttbar{}{5}{5}
              \ganttbar{}{9}{9}
              \ganttbar{}{13}{13}
              \ganttbar{}{17}{17}
              \ganttbar{}{21}{21}
              \ganttbar{}{25}{25}
              \ganttbar{}{29}{29}
              \\
              \ganttbar[bar/.append style={fill=orange!50}]{auxiliary thread}{10}{12}
              \ganttbar[bar/.append style={fill=black!10}]{}{13}{13}
              \ganttbar[bar/.append style={fill=orange!50}]{}{14}{15}

              \ganttbar[bar/.append style={fill=orange!50}]{}{18}{20}
              \ganttbar[bar/.append style={fill=black!10}]{}{21}{21}
              \ganttbar[bar/.append style={fill=orange!50}]{}{22}{23}

              \ganttbar[bar/.append style={fill=orange!50}]{}{26}{28}
              \ganttbar[bar/.append style={fill=black!10}]{}{29}{29}
              \ganttbar[bar/.append style={fill=orange!50}]{}{30}{31}

       \end{ganttchart}
       \caption{Multi-threading. The audio thread (highest priority) and the auxiliary thread are shown, respectively, in red and orange. Grey indicates that the thread is sleeping. In this example, the auxiliary thread waits for a buffer of two blocks of samples to be filled. For this reason, the auxiliary thread does not run in the first two blocks.}\label{fig:mthread}
\end{figure*}

The audio callback (in the Bela API, the \texttt{render()} function) is a function in the audio thread that is called for each block of samples and does the audio processing. In Figure~\ref{fig:mthread}, this is indicated by the red boxes. For simplicity, we assume that a block of samples processing always takes the same amount of time (a fourth of a block) and that the inference task only takes a block and a fourth\footnote{\pnew{In reality, the inference task, due to its computational complexity, would probably take longer (e.g. 16 audio blocks).}}. \pnew{If the inference task is called from the audio thread it will not be able to finish on a block's time, which will cause an underrun. Increasing the block size would ensure that the inference computation finishes on time, however, large block sizes add significant inherent latency to the system. }


\pnew{Tasks that are executed occasionally} (i.e. not in every block, for example, after filling an input buffer) and that are computationally expensive\pnew{, such as the inference task,} should be called from an auxiliary thread with a lower priority than the audio thread. This is shown in Figure~\ref{fig:mthread}: in orange, the inference task is called when the input buffer is filled with two audio blocks (i.e. at the end of block 2). However, it does not start running immediately at the beginning of block 3, since the audio callback has higher priority. Once the audio callback task is finished, the inference task starts executing. When block 4 starts, the inference task is put to sleep (indicated with a gray bar), and the CPU executes the audio callback instead. When the audio callback task is finished, the CPU is free again to complete the inference computation. At the end of block 4, the input buffer has been filled again (with samples from blocks 3 and 4), and the process repeats.



\section{Conclusion}
Many embedded platforms have lowered entry barriers to real-time audio programming by abstracting complex tasks through APIs \pnew{and IDEs}. However, deploying deep learning models involves an ecosystem of tools that demands lower-level software development skills, such as building a program using a custom CMake recipe. This paper presents a pipeline for recording datasets and deploying neural models in the Bela embedded hardware platform. In contrast to existing embedded AI projects that target a specific deep learning task or application, this pipeline serves as a template \pnew{for programmers and makers to prototype and experiment with deep neural networks for real-time musical applications}. With this pipeline, we aim to reduce the complexity (or viscosity \cite{Blackwell2003}) of this process and contribute towards bridging the deep learning for audio and embedded hardware communities in NIME.

As part of the pipeline, we provide tools for recording multichannel datasets (by connecting multiple Bela \pnew{boards}) and a dockerised cross-compilation environment. These tools are of interest beyond their role in the pipeline: the dataset recording tool allows capturing datasets from multichannel sensor arrays for later analysis, and \pnew{the cross-compiling environment significantly reduces the compilation times}.

Finally, the models will only run in real-time if their computational complexity is low enough to run on the Bela CPU. Future work will focus on including \pnew{network compression} and embeddability diagnosing tools into the pipeline.

\section{Acknowledgments}
\pnew{This work was supported by the EPSRC UKRI Centre
       for Doctoral Training in Artificial Intelligence and Music (EP/S022694/1) and the Royal Academy of Engineering under the Research Chairs and Senior Research Fellowships scheme.
       The authors would like to thank Giulio Moro and Franco Caspe for their technical support, Lia Mice for the mini Chaos Bells lending, and the anonymous reviewers for their valuable feedback. }

\section{Ethical Standards}
\pnew{Adán L. Benito and Andrew McPherson are part of Augmented Instruments Ltd, the company that produces the Bela platform. Teresa Pelinski's PhD is also partly supported by the same company. }

\bibliographystyle{abbrv}

\bibliography{nime.bib} 


\end{document}